\documentclass[aps,preprint,preprintnumbers,amsmath,amssymb, floatfix]{revtex4}
\usepackage{bm}
\def\be{\begin{equation}}
\def\ee{\end{equation}}
\def\beq{\begin{eqnarray}}
\def\eeq{\end{eqnarray}}

\begin{document} \openup8pt 
\preprint{smw-genart-02-06}
\title{Beyond Einstein ... Are we all afraid of the Truth?}

\author{Sanjay M Wagh}

\affiliation{Central India Research Institute, Post Box 606,
Laxminagar, Nagpur 440 022, India  \vspace{1in}}

\begin{abstract}
The power-point presentation \cite{ppt} provided herein shows
exactly why Einstein's field equations of his general relativity
are based on an illogical approach to representing the observable
world. Einstein had, in fact, discarded these equations way back
in 1928 when he had began his solitary search for a unified field
theory. However, the rest of us learned, taught, and also put too
much faith for too long (for more than seventy years) in an
illogical approach to representing the observable world.
Consequently, we have developed great reluctance, resulting from
dogmatic perceptions, prestige, reputation, ..., that is holding
us back from orienting ourselves in the ``right'' direction to the
understanding of the observable phenomena. This raises the
question mentioned in the title: Are we all afraid of the Truth?
Rhetorically speaking, we could then also ask: are we all afraid
of Virginia Woolf? In the sequel, I also illustrate my approach to
going Beyond Einstein for developing an appropriate mathematical
framework for the fundamental physical ideas behind the General
Principle of Relativity, for the unification of fundamental
physical interactions and, hence, for a theory of everything.
\end{abstract}

\email{cirinag_ngp@sancharnet.in}

\date{March 20, 2006}
\maketitle

\newpage
The title of this article may appear only as an eye-catching one
to some, repelling one to some others, a thought provoking one to
few others ... Only the person reading it can tell. However,
before we plunge ourselves into the issues related to this title,
a preamble to it appears necessary to setup the backdrop against
which it should be viewed to get an insight into the question it
poses.

Physics is our attempt to conceptually grasp the happenings of the
observable world. Concepts of Physics are also succinctly
expressible in the language of mathematics. Then, to obtain a
genuine understanding of the physical phenomena, observable
changes in physical or observable bodies, it is necessary for us
to formulate appropriate concepts and to also express them by
proper mathematical notions.

This last issue is clearly perceptible \cite{laue} with the
pre-Newtonian development of Mechanics. The concept of the motion
of a physical body involves change in its ``location'' relative to
a reference body - another physical body. This ``location'' is
physically ``measurable'' using still another, third, physical
body as a measuring scale. This measurement yields a ``number'' -
the ``distance'' between the reference body and the physical body
whose ``location'' is being ascertained with respect to the
former. The motion of the first physical body is then
mathematically expressible as a rate of change of this number,
distance, with respect to Time. [Time here is a concept related to
the ``location'' of the ``hand'' of a clock body relative to the
reference physical body. Here, we will not enter into further
details about Time, not that these are unimportant.]

Because the distance is a pure number, a scalar, its rate of
change with time is a scalar, the speed. Descartes then used this
notion of speed to formulate and propose the Laws of Motion of
observable bodies using it.

Observations with collisions of physical bodies did not, however,
support the Laws of Motion as proposed by Descartes.

Huygens then realized that if we appropriately associated a
positive and a negative sign with the speeds of two colliding
bodies, then the Laws of Motion so obtainable are in agreement
with the observed collisions of bodies. This assignment of
positive or negative sign to the speed was only an ad-hoc proposal
of Huygens and was restricted only to his considerations of the
collision of two physical bodies.

Of course, it was yet to be realized that an appropriate notion to
describe motion is that of ``displacement'' and that it involves
not just the change in distance to a physical body but also the
``direction'' of this change in its distance. That is to say, it
was yet to be realized that we have to treat the ``displacement''
as a vector.

As is well known, Newton, with a deeper insight than his
contemporaries, provided us a logically consistent conceptual
framework for Mechanics using vectors.  As Newton's Laws of Motion
were in ``agreement'' with most subsequent experiments, the
concept of motion then found its proper mathematical
representation. Descartes and others also provided geometrical
background to these notions by invoking the Euclidean geometry and
the mathematical notion of a (Cartesian) coordinate system. This
is the representation of physical bodies by a point of the
Euclidean 3-dimensional space.

Experiments ``verify'' theoretical explanations of phenomena and,
in turn, indicate the appropriateness of our choice of, both, the
physical conceptions and the mathematical structures representing
them. Due mainly to Galileo, experimental methods of determining
this appropriateness were already well established before Newton
who always used such methods to substantiate his theoretical
conclusions.

To state his Laws of Motion, Newton introduced an important
concept, Force, as a ``cause'' behind the motion. He defined force
as the rate of change of (Descartes's) quantity of motion, the
momentum vector, with time. He then also related force with the
acceleration and the (Galilean) inertia of a physical body.

There also are purely logical methods to decide, at least partly,
the appropriateness of concepts. These determine the {\em mutual
compatibility\/} of our concepts, {\em ie}, the {\em internal
consistency\/} of the theoretical framework. Newton also used
logical methods, {\it eg}, the formulation of his third law of
motion, to ``construct'' theory.

As only a physical body should be the ``cause'' behind a physical
phenomenon, every force was expected to have physical bodies as
source(s). This, however, was not to be with Newton's definition
of force. The {\em Coriolis force\/} did not have any source in
physical bodies. This {\em pseudo-force\/} only arises because the
chosen reference body, the Earth, is rotating relative to some
``standard'' reference physical bodies.

What then are these ``standard'' reference bodies? To analyze this
issue, Newton devised the famous bucket experiment. This thought
experiment was an indication to Newton himself that his conceptual
framework was not logically completely satisfactory. There also
were observations from Optics to indicate to Newton that his
framework of Mechanics was insufficient to explain all the
physical phenomena.

Moreover, Newton had mathematically represented a physical body as
a point of the underlying 3-dimensional Euclidean space. Newton's
laws of motion were then closely related to certain mathematical
properties of (time-parametrized) curves of the Euclidean
3-dimensional geometry, and, as is well known, Newton himself
developed corresponding mathematical notions of Calculus.

In this connection, Descartes \cite{schlipp, ein-pop} rightly
pointed out that this representation of a physical body by a point
of the Euclidean space is not in {\em conformity\/} with our
everyday experience that reference bodies do get affected by
physical processes. Specifically, he was concerned about the fact
that the coordinate system of the Euclidean 3-dimensional space
underlying Newtonian Mechanics, when viewed as a ``material
construction'' of a reference body, does not get affected by
physical precesses.

Generically, using checks of either experimental origin or logical
origin (the internal consistency of the theory), we judiciously
accept or reject any conceptual framework as an admissible theory
of the observable world. When an internally consistent theory
fails to explain some observations, we need to {\em expand\/} the
conceptual basis of that theory and, hence, mathematical
structures representing those concepts. As an acceptable
explanation of the observable world, the conceptual framework of
the ``expanded'' theory must also be internally consistent in the
sense of Logic.

But, alternative mathematical notions were simply not available in
Newton's times. Therefore, although Newton and a few others
``sensed'' that the mechanistic framework needed ``modifications''
at fundamental levels, neither Newton nor could anyone else (of
Newton's times, sensing this need) suggest any alternative to it.

The world, then, got blinded by successes of the Newtonian Theory,
and fell pray to the Mechanistic Dogma  - that the entirety of
physical phenomena could be explained using the Newtonian
conceptions from Mechanics.

Scientific developments after Newton followed the path of
``experimental'' checks. Mechanistic Dogma and the lack of
alternative mathematical notions, both, prevented the majority
from exploring alternatives to Newton's Mechanics.

As is well known, only the experimental data, increasingly getting
inconsistent with Newtonian theoretical predictions, ultimately
severed, once for all, strong links with the Mechanistic Dogma of
the post-Newtonian era.

Perhaps, if fundamental limitations of Newton's theory were widely
known, mechanistic dogma would not have gripped the majority in
the first place. Of course, the lack of means of rapid
communication was the reason for these limitations not being
widely known. Only a few centuries after Galileo and Newton,
Einstein could free us, still only partially \footnote{Even during
Einstein's times, the rapid means of communication were
unavailable. Access to needed information in printed research
journals was severely limited and involved inordinate delays.
Media provided scientific news from time to time. Media therefore
played an exceptionally important role in educating the masses
then, and continue to play a similar role even in the modern
IT-era.}, from the dogmas associated with Newton's theoretical
constructions.

I am referring here to Einstein's famous contributions
\cite{schlipp} to two well known revolutions in Physics, {\em
viz}, Special Theory of Relativity and Quantum Theory. Einstein's
analysis of the role of Light in physical measurements freed us
from ``fixed'' Euclidean 3-space and ``universal'' time, both of
the Newtonian era. On the other hand, Schr\"{o}dinger's and
Heisenberg's mathematical methods of Quantum Theory \cite{std-qm}
showed us, among other things, that mathematical methods
significantly different than those of the Newtonian theory do
describe the observable world in a better manner.

With these revolutions, Science, in general, has developed so
rapidly that it has greatly influenced almost every aspect of our
lives. In comparison to Newton's and Einstein's eras, we have
rapid, almost instantaneous, means of communication, thanks to the
technological revolution with Computers, the Internet... . Then,
with these developments, experimental methods have also advanced
and have also become exceptionally cutting into the pockets of the
tax payers globe over. The role of ``all of us'' in promoting an
open dialogue over fundamental physical conceptions and their
limitations is therefore quite important to free us from
associated dogmatic perceptions, if any.

The above appears to be presently very relevant simply because
dogmatic perceptions do appear to prevail within the physics
community at large. The following discussion considers these
dogmatic perceptions from my perspective.

First such deep-rooted dogma appears to be that mathematical
methods of Quantum Theory will \cite{string, aa-lqg} lead us to a
{\em Theory of Everything}. Here, it is then forgotten that
mathematical methods of Quantum Theory do not represent physical
bodies in such a way that the effects of physical phenomena on any
reference body are encompassed by this mathematical
representation. That is to say, Descartes's criticism of Newton's
mathematical representation of a physical body as a point of the
Euclidean space applies also to this quantum theoretic
representation of a physical body.

The issue here is also that of the limitations of the conceptual
framework as well as those of the mathematical methods of Quantum
Theory as discussed in \cite{smw-utr}. Specifically, quantum
theory uses inertia, potential, source of the potential and a law
of motion as fundamental notions. These four notions are however
{\em mutually logically independent}. Hence, no explanations of
any kind for these four basic notions are possible within the
conceptual framework of the quantum theory.

The mathematical framework of the quantum theory faithfully
expresses its physical conceptions and is, consequently,
inadequate to remove the mutual logical independencies of its
above basic notions. That is to say, the (extended) Lagrangian, or
the Hamiltonian, method needs the aforementioned four conceptions
to first yield mathematical expressions to correspond to them, and
only after this has been done that the method provides the
evolution of the system that it represents. But, neither that
Lagrangian nor that Hamiltonian is obtainable within this
framework.

Then, the mathematical framework of the quantum theory can do no
better than the (extended) lagrangian or hamiltonian frameworks,
which cannot of course remove the mutual logical independencies of
the aforementioned four basic notions of the quantum theory. A
question is also whether these frameworks can, with {\em prior\/}
specifications of {\em all\/} the relevant physical conceptions by
appropriate mathematical expressions, account for the entirety of
physical phenomena.

This is really an issue of non-lagrangian and non-hamiltonian
mathematical systems. The point is that mathematical such systems
do exist. Then, if a theory better than the quantum theory were to
be fundamentally based on such mathematical notions, lagrangian or
hamiltonian representation for every physical phenomenon may not
exist within it. A better theory here must first remove the mutual
logical independencies of the fundamental physical concepts
mentioned before.

That is why we need to ``look beyond'' the mathematical framework
of the Quantum Theory for developing an appropriate theoretical
framework within which the mutual logical independencies of
fundamental physical notions are removed \cite{smw-utr}. Such a
``framework'' may be termed as a {\em Theory of Everything}. We
already have a ``good'' understanding of physical phenomena of the
microscopic domain (corresponding to electromagnetic, weak and
strong nuclear interactions) on the basis of the quantum theory
and an appropriate inclusion of gravitational phenomena within
this ``framework'' may, thinkably, exhaust the entire list of
observable phenomena needing description.

But, many may have developed a frame of mind that such a theory of
everything is not possible. Surely, we have to be cautious with
any such claim, but to turn completely away from the possibility
of a Theory of Everything is dogmatic.

In this connection, we recall that the purpose of physical science
is to ``describe'' changes in physical bodies as are observed.
That such a description is possible is the basis of our developing
physical theories.

As we have stressed before, mathematics is the language of
physical science and physical concepts are also succinctly
expressible using mathematical notions. Then, we may think of a
{\em single\/} mathematical notion to ``represent'' not only {\em
all\/} the characteristics of physical bodies but also their
``changes'' (mathematical transformations).

Here, we then note that a mathematical transformation essentially
``knows'' about the mathematical structure it ``transforms''. This
single concept, that of the transformation of a mathematical
structure representing {\em all\/} the characteristics of physical
bodies, appears to possess therefore the ingredients necessary to
be the single conceptual entity that may be the basis of a Theory
of Everything. That this description may really be possible
\cite{smw-let-0601} goes against the dogma that a Theory of
Everything is impossible.

Another dogmatic perception appears to be that Einstein's field
equations of General Relativity should describe well the
phenomenon of gravitation at least in some suitable approximation
to a better theory. Consequently, predictions obtainable from the
analysis of these equations are considered \cite{maccallum} as
{\em physically relevant}.

With Schwarzschild's monumental discovery in 1916 of a spherically
symmetric solution to these highly non-linear, partial
differential equations, the attention of the world, in particular,
of various mathematicians, had turned to these equations
\footnote{Of course, David Hilbert, a mathematician, had also
proposed these equations.}. Important mathematical methods related
to solving non-linear partial differential equations were
developed and numerous solutions \footnote{However, it should also
be noted here that various properties of many of these solutions
are, from the physical point of view, exceedingly perplexing, for
example those of the Taub-NUT and other spacetimes.} of Einstein's
field equations were obtained. This is still an active area of
research in General Relativity.

In recent times, to be specific since 1963, the discovery of Quasi
Stellar Objects and high energy phenomena in Active Galactic
Nuclei have brought into prominence with physicists and
astrophysicists these solutions of Einstein's field equations and
methods of perturbational handling of these equations themselves.
The Black Hole solutions of these equations are also considered
\cite{chandra} to be very significant advances in understanding
the related astrophysical events, considering particularly that
precursors to Black Holes, the Neutron Stars, have been discovered
in the form of Pulsars.

Interesting speculations about possible detection of the Black
Hole or the Event Horizon in astronomical observations of certain
X-ray sources have also been attempted \cite{bhsupport}. These
have also found places as headlines with various media. In a
``scientifically aware society'', this is certainly desirable as
well as inevitable.

The same applies to astrophysical models of the entire Universe -
the Cosmology. Hubble observed that the red-shifts of distant
galaxies increase with their distance from us. When red-shift is
interpreted \footnote{But, is this interpretation of red-shift
using Doppler's effect correct? Is there another permissible
explanation for (part of) the red-shift that is more significant
in the cosmological context? These are still open issues.} using
Doppler's effect, this all leads to an interesting picture of an
expanding universe of galaxies. One implication of this could be
that galaxies had been closer to each other in the past.

Then, the Universe of Galaxies should have ``originated'' out of a
{\em unique\/} event in the history of this Universe. This is then
as if matter had been thrown out in a Big Explosion - the Big
Bang. The explosive event imagined here should have associated
with it radiation, which should decouple from matter at some stage
because of cooling due to expansion, and should be observable as a
relic of the Big Bang. The observed Cosmic Microwave Background
Radiation is then ``naturally'' explainable on the basis of the
Big Bang assumption as its relic radiation.

Apart from the Big Bang conception, an alternative explanation of
the expansion of the Universe of Galaxies is also possible if, in
addition to the attractive force of gravity, some suitable
``repulsive force'' between galaxies is assumed. With an
appropriate behavior of this repulsive force, galaxies then need
not originate out of any unique event such as a Big Bang. This
explanation also becomes as attractive as the Big Bang proposal
when ``natural'' explanation is provided for the existence of this
repulsive force. This is the basis of the Steady State Cosmology
\cite{jvn01, fhjvn}.

The sixties had thus witnessed the famous debate of Big Bang
versus Steady State Cosmology. It was then subsequently claimed
that the Steady State Model is untenable vis-a'-vis observations
of radio sources. So much had been the general, justifiable,
interest in these Cosmological issues that this scientific debate
also found an appropriate place in various enjoyable novels of
satirical nature!

But, a cursory glance at the submissions, of theoretical as well
as observational nature, to various databases is sufficient to
show that this debate is not settled as yet. However, the
``majority'' seems to be favoring the Big Bang Model and most
researchers therefore interpret observations from only the point
of view of this model.

As observations improved, the need to postulate the aforementioned
repulsive force has become more imminent. Observations appear to
indicate that the rate of expansion of the universe is currently
increasing. Then, the Big Bang conception needs to explain these
observations. Any explanation is possible only if some repulsive
force has, in the recent times, become ``operative'' to accelerate
the expansion. Within this scenario, Inflation, Quintessence,
Coasting etc.\ are just manifestations of different behaviors of
the assumed repulsive force, needing natural origin \cite{jvn01}.

A preliminary mathematical rendering of this physical picture of
the universe at large was then to be ``found'' within the
homogeneous and isotropic Friedmann-Lemaitre-Robertson-Walker
geometry \cite{jvn01} for the spacetime. Of course, this
geometrical rendering is {\em valid\/} if the galaxies are
homogeneously and isotropically distributed.

The corresponding solution of the Einstein field equations then
represents the Big Bang picture. In this case, the ``initial
singularity'' of the FLRW spacetime then signifies the unique
event of the Big Bang. But, the spacetime singularity implies the
breakdown of the geometric description in terms of Einstein's
equations. The problem of the spacetime singularity, it was hoped,
would then go away with the ``quantum'' considerations related to
geometry, {\em ie}, with the Quantum Theory of Geometry
\cite{aa-lqg, qg-bang}.

On the other hand, the Steady State Cosmology uses \cite{jvn01}
the same homogeneous and isotropic FLRW spacetime geometry, but
its equations of evolution are modified away from those of
Einstein's theory of gravity by the presence of the creation-field
terms producing the repulsive force mentioned before. The latest
version of this theory, the Quasi Steady State Cosmology, then
assumes \cite{jvn01} certain behavior for the creation-field
terms, but the original philosophical appeal is then lost for
many.

But, it is crucial to realize that Einstein's equations are, as
the provided power-point presentation clearly shows, based on an
illogical and, hence, unscientific, approach to explaining the
physical world.

Even if these equations were assumed to provide some ``geometric''
explanations of observable phenomena, the corresponding conceptual
formalism \footnote{Einstein's intentions behind these equations
was that of the representation of the entire physical reality. See
later.} is not satisfactory. This is because electromagnetic
phenomena will then not be geometrically explainable as arising
due to the curvature of the underlying geometry.

We also note here that Einstein had, in fact, begun discarding
these equations \cite{ein-pop, schlipp, pais} by 1928 when he
started on his solitary search for some satisfactory Unified Field
Theory of gravity and electromagnetism.

Moreover, the action-at-a-distance formalism cannot remove
\cite{smw-utr} the mutual logical independencies of the
fundamental physical conceptions mentioned earlier. This is so
because the action-at-a-distance framework, unavoidably, needs to
associate ``source characteristics'' with observable bodies
corresponding to assumed forces. Explanations for such source
characteristics and the assumed Law of Force are then outside the
scope of any action-at-a-distance framework as was Newton's.

This is what brings us to the question raised in the title of this
article. It refers to above situations with the aforementioned
dogmatic perceptions associated with the two fundamental pillars
of the modern physical science.

In what follows, we will discuss some lopsided developments of
ideas when alternative explanations to concerned phenomena should
really have been explored. These lopsided developments appear to
be based on the aforementioned dogmatic perceptions and, hence, on
inappropriate conceptions. Unjustifiably, alternative explorations
were, time and again, simply shunned as a result of dogmatic
perceptions.

Due to many dogmatic perceptions, there has only been a lopsided
development of models of the Big Bang conception, and that too
using the inappropriate mathematical framework of the Einstein
field equations of general relativity. Observations have mostly
been interpreted from the point of view of this conception.
Alternative cosmological scenarios have not been explored to the
same level of details and researchers exploring such alternatives
do not find the required support \footnote{Substantial support
comes neither with the refereed journals nor with the funding
agencies. Most journals then shoot a stereotyped rejection or the
referees reject such speculative research without paying much
attention to alternative ideas.}.

This lopsided development is also seen with the models of
astrophysical bodies such as Quasars and Active Galactic Nuclei
based on an entirely inappropriate notion of general relativistic
Black Holes, not worth mentioning here are the naked singularities
(that, justifiably, were complete anathema to Einstein).
Alternative models have also been explored, no doubt, but only by
the minority of researchers who find it hard to garner support for
their results. An already famous such case related to alternative
ideas has been that of H Arp's Quasar-Galaxy associations
\cite{jvn01}.

Such situations have arisen because of dogmatic perceptions at
various levels of the scientific echelons. If not, then how else
do we explain these lopsided and inappropriate developments of
only certain ideas? One could of course blame the lack of means of
rapid communication in the past for the propagation of various
inappropriate conceptions. But, this alone does not completely
absolve us of the relevant sin.

In order to proceed ``Beyond Einstein,'' let us then recall at
this place Einstein's personal approach to his own theoretical
constructions. The pivotal point of Einstein's formulation of
relevant ideas is the equivalence of the {\em inertia\/} and the
{\em gravitational\/} mass of a physical body, a fact known since
Newton's times but which remained only an assumption of Newton's
theory.

On the basis of this {\em equivalence principle}, Einstein then
arrived at the {\em General Principle of Relativity\/} that the
Laws of Physics be applicable with respect to \underline{all} the
systems of reference, in relative acceleration or not, {\em
without unnatural forces} (whose origin is not in physical or
observable bodies) {\em entering into them}. Then, the Laws of
Physics should be based on the same mathematical structures, and
be also the same mathematical statements, for all the reference
systems.

The {\em equivalence principle\/} implies that the Lorentz
transformations of the Special Theory of Relativity are not
sufficient to incorporate the explanation of this equivalence of
inertia and the gravitational mass of a material body.
Furthermore, it also follows that general transformations (of
coordinates) are required by the equivalence principle, and the
physical basis is then that of the {\em general principle of
relativity}, which is certainly more appealing than the restricted
special principle of relativity.

To arrive at his formulation of the general theory of relativity,
Einstein first raised \cite{schlipp} (p. 69) the following two
questions:

\begin{itemize} \item {\em Of which mathematical type are variables (functions
of coordinates) which permit expression of physical properties of
space (``structure'')?} \item {\em Only after that: Which
equations are satisfied by those variables?}
\end{itemize}

In 1949, more than thirty years after he proposed his field
equations of the general theory of relativity in 1916, he still
wrote \cite{schlipp} that:

\begin{quotation} {\em The answer to these questions is today by no means
certain.} \end{quotation}

He then elucidated his steps to these questions by his
considerations of the
\begin{description} {\em \item{(a)} pure gravitational field
\item{(b)} general field (in which quantities corresponding
somehow to the electromagnetic field occur, too).}
\end{description}

He then wrote \cite{schlipp} about his approach to a general
theory of relativity in the following ``recollective'' words that:

\begin{quotation} {\em It seemed hopeless to me at that time to venture the attempt
of representing the total field (b) and to ascertain field-laws
for it. I preferred, therefore, to set up a preliminary formal
frame for the representation of the entire physical reality; this
was necessary in order to be able to investigate, at least
preliminarily, the usefulness of the basic ideas of general
relativity.} \end{quotation}

In 1916, as is well known \cite{maccallum, jvn01}, he then
proposed the following equations for his p\underline{reliminar}y
explorations of the representation of the \underline{entire}
p\underline{h}y\underline{sical} \underline{realit}y:
\[ R_{_{ij}}-\frac{1}{2}\,R\,g_{_{ij}}=-\,\kappa\, T_{_{ij}}\]
where the left hand side is a geometric quantity while the right
hand side is the energy-momentum tensor representing physical
matter. These equations reduce to appropriate Newtonian equations
in a suitable limit.

In 1949, he still expressed his concerns \cite{schlipp} about
these preliminary equations in the following words:

\begin{quotation} {\em The right side is a formal condensation of all things whose
comprehension in the sense of a field theory is still problematic.
Not for a moment, of course, did I doubt that this formulation was
merely a makeshift in order to give the general principle of
relativity a preliminary closed expression. For it was essentially
not anything {\em more\/} than a theory of the gravitational
field, which was somewhat artificially isolated from a total field
of as yet unknown structure}. \end{quotation}

In 1949 again, he further wrote \cite{schlipp} about his
perception of any satisfactory formulation of a theory
incorporating the general principle of relativity in the following
words clearly indicating that ``geometric'' singularities are
anathema (my underlining):

\begin{quotation} {\em Maxwell's theory of the electric field remained a torso,
because it was unable to set up laws for the behavior of electric
density, without which there can, of course, be no such thing as
an electromagnetic field. Analogously the general theory of
relativity furnished then a field theory of gravitation, but no
theory of the field-creating masses. (\underline{These}
\underline{remarks} p\underline{resu}pp\underline{ose}
\underline{it} \underline{as} \underline{sel}f-\underline{evident}
\underline{that} \underline{a}
f\underline{ield}-\underline{theor}y \underline{ma}y
\underline{not} \underline{contain} \underline{an}y
\underline{sin}g\underline{ularities}, i.e., \underline{an}y
p\underline{ositions} \underline{or} p\underline{arts}
\underline{in} \underline{s}p\underline{ace} \underline{in}
\underline{which} \underline{the}
f\underline{ield}-\underline{laws} \underline{are} \underline{not}
\underline{valid}.)} \end{quotation}

Einstein had, in 1928, concluded that his equations of General
Relativity were not any satisfactory formulation of the physical
reality. He had attempted numerous formulations for the unified
field theory. Pauli \cite{pais, pauli} (p. 347 of \cite{pais})
then criticized: \begin{quotation} {\it [Einstein's] never-failing
inventiveness as well as his tenacious energy in the pursuit of
[unification] guarantees us in recent years, on the average, one
theory per annum ... It is psychologically interesting that for
some time the current theory is usually considered by its author
to be the ``definitive solution''...}
\end{quotation} and had already demanded to know from Einstein, in
a letter to Einstein dated December 19 1929, as to ``what had
become of the perihelion of Mercury, the bending of light, and the
conservation laws of energy-momentum.''

Einstein had no good answers to these questions within his Unified
Field Theory. He however was not overly concerned about the issues
raised by Pauli. But, in a letter written on January 1, 1930 to W
Mayer he wrote: (my underlining)
\begin{quotation} {\it Nearly all the colleagues react sourly to
the theory because it p\underline{uts}
\underline{a}g\underline{ain} \underline{in} \underline{doubt}
\underline{the} \underline{earlier} g\underline{eneral}
\underline{relativit}y.}
\end{quotation} In 1931, in a note \cite{ein-sci} he also admitted
that his earlier attempts at a unified field theory constituted a
{\em wrong direction\/} to follow.

Interestingly, in the early forties, Einstein also explored
\cite{pais} (p. 347) the question of whether the most fundamental
equations of physics might posses mathematical structure other
than that of the partial differential equations.

{\em As Einstein's field equations are based on logically
unacceptable substitution of only the force of gravity by the
curvature of the spacetime geometry, their solutions cannot form
any logically or scientifically acceptable explanations of
observable phenomena. Hence, explanations based on black hole
solutions, the naked singularity solutions as well as the analysis
of cosmological solutions of these equations, including
perturbations, are no explanations of the concerned physical
phenomena.}

Einstein had very clearly recognized that a field theory must not
contain any singularities as particles. However, perhaps under the
pressure of the ``Publish or Perish'' syndrome and due to lack of
relevant information, both, we attempted unscientific descriptions
of physical phenomena for over the last century almost. We also
gave to the general masses an impression of satisfaction with such
descriptions.

The Truth however remains that Einstein's field equations are
logically unacceptable as forming any basis for scientific
explanations of the observable world. Origins of the persistent
dogmatic perception, that Einstein's equations describe the
behavior of matter when gravity dominates, rest then with the
following situations.

It is certainly true that Einstein's equations provide us a
mathematical method of determining a 4-dimensional geometry.
Properties of the 4-dimensional geometries are of course based on
mathematically consistent methods.

Therefore, equations of geodesics of the geometry, equations of
geodesic deviations, perturbational analysis of the underlying
geometry, etc.\ are mathematically consistent. It also is quite
justifiable that we construct \cite{aa-lqg} 4-dimensional
geometries using ``non-metric building blocks made up of'' spinor
variables.

Consequently, any use \cite{aa-lqg} of quantum theoretic methods
for the Einstein-Ashtekar gravity can also be expected to provide
us some mathematically consistent framework of the ``Quantum
Theory of Geometry''.

But, how can these methods yield us any acceptable \footnote{Some
beautiful part of Mathematics may not have anything to do with
observable situations.} explanations of the observable phenomena
if their basis, Einstein's equations themselves, is logically
unacceptable? How do the conceptual difficulties \cite{qgproblems}
with the quantum theory of the spacetime get resolved? These
methods are also based on the lagrangian or hamiltonian formalism.
In the absence of logically consistent underlying conceptions of
Physics, what have these mathematical methods got to do with the
observable phenomena?

The Truth also remains that procedures of quantum theoretic origin
as used in highly innovative approaches to unification of basic
forces are based on lagrangian or hamiltonian methods, which
cannot provide explanations for the aforementioned four
fundamental conceptions of the quantum theory. No doubt, the
conceptions behind various attempts such as superstring theory,
Euclidean quantization program etc.\ have been ingenious and
highly innovative. The fact still remains that these approaches
are explicitly or implicitly based on lagrangian or hamiltonian
method.

How then can the unification of fundamental interactions be
possible in any of such approaches \cite{string}? We could again
ask, what happens to conceptual difficulties with the quantum
spacetime here? How do these difficulties get resolved?

Is not an open-minded recognition of such facts by all of us
required? We should remember what Einstein had said:
\begin{quotation} {\em Anyone who has never made a mistake has never
tried anything new.}
\end{quotation} How true this is! As a child, we all make a mistake,
learn from that mistake and turn away from that mistake to other
matters. Is that why we, as children, are creative, innovative,
and always exploring something new?

The elders are reluctant to admitting a mistake, perhaps because
of their false notions of pride, reputation ... Should we be prey
to such false notions? Then, is it not in the wider scientific
interests to accept (in reality, Einstein's) mistake and progress
along the right path to a better understanding of the observable
world?

Einstein was an honest man. Unperturbed by his failures, he had
openly admitted to his formulation of a theory of general
relativity in terms of his equations being not right. Unafraid of
losing his reputation, he had solitarily pursued the unified field
theory. Again, he had openly admitted to his path being not the
right one to follow. Due only to his such qualities, a common
person regards him as an Icon of Science.

The then torch-bearer of physical science, Einstein, had shown us
that the path of his field equations of general relativity is an
incorrect one to follow. Due however to the lack of relevant
information in the past, we ``followed'' that incorrect path even
to this date. In the present IT-era, we are fortunate to possess
the means of rapid communication. Then, are we to discard the
inappropriate path and ``search'' for the ``correct'' path to the
understanding of the physical world?

My purpose behind this article is to criticize (but not in any
manner belittle other lines of thoughts) and to plainly state all
the related facts so that we could all rethink about pros and cons
of approaches to explaining the observable world.

History of Science beyond Galileo and Newton has shown that only
right ideas survive and the incorrect or inappropriate ones simply
vanish into oblivion. To accept a mistake is the right thing to
do, and Einstein had openly admitted to that mistake. It is then a
right way to go Beyond Einstein. This is a way to genuine
progress. Has not the Time come today for the majority of us to
take a definitive stance in favor of the Truth and to take steps
in the ``right'' direction?

Are we then progressive? Or, are we all afraid of the Truth? And,
because we are afraid of the Truth, shall we suppress and not let
the Truth be out? Then, rhetorically speaking, are we all afraid
of Virginia Woolf?

\acknowledgments I am grateful to my brother Hemant Wagh and to
Dilip A Deshpande  for proposing that I write this article, for
reading it, for suggesting improvements to it and also for various
discussions about concerned issues from time to time.


\newpage
\begin{thebibliography}{99}
\bibitem{ppt} Wagh S M (2006) file:{\bf no-efogr.ppt}. This is a
part of the talk {\em How do we understand the Universe around
us?\/} given on the occasion of the World Year of Physics 2005 at
the Janakidevi Bajaj Science College, Wardha on January 7, 2006,
and also as a part of the talk {\em Information Communication
Technology in Education\/} to participants of the Seminar on ICT
in Education organized by the Institution of Electronics \&
Telecommunication Engineers, Nagpur Center and the VMV Commerce,
JMT Arts \& JJP Science College, Wardhaman Nagar, Nagpur on
February 4, 2006, and also as a part of the talk {\em Use of
Information Communication Technology in Teaching and Education\/}
to science and non-science College Teachers at the Academic Staff
College, Nagpur University, Nagpur during an Orientation Program
on March 6, 2006

\bibitem{laue} See an excellent article by von Laue M (1970)
in {\it Albert Einstein: Philosopher-Scientist}\, (Ed. P A
Schlipp, La Salle: Open Court Publishing Company - The Library of
Living Philosophers, Vol. VII).

\bibitem{schlipp} Einstein A (1970) in {\it Albert Einstein:
Philosopher Scientist} (Ed. P A Schlipp, Open Court Publishing
Company - The Library of Living Philosophers, Vol VII, La Salle).
See, in particular, his Autobiographical Sketch and his reply to
essays by others.

\bibitem{ein-pop} See, Einstein A (1968) {\it Relativity:
The Special and the General Theory\/} (Methuen \& Co. Ltd, London)
(Appendix V: Relativity and the Problem of Space.)

\bibitem{std-qm} See any of the many standard books on Quantum Theory. For
example, Varadrajan V S (1988) {\it Geometry of quantum theory},
Vol. I and II (Van Nostrand Reinhold Co., New York)\\ Dirac P A M
(1970) {\it Principles of Quantum Theory\/} (Dover, New York) \\
Jauch J M (1968) {\it Foundations of Quantum Mechanics},
(Addison-Wesley, Reading)  \\ Also, Messiah A M L (1960) {\it
Quantum Mechanics}, (North-Holland, Amsterdam) \\ Richtmyer F K \&
Kennard E H (1942) {\it Introduction to Modern Physics}, (MacGraw
Hill, New York)

\bibitem{string} Greene M B, Schwarz J H \& Witten E (1987) {\em
Superstring Theory}, Vol I and II (Cambridge University Press,
Cambridge)  \\ Polchinski J (1998) {\em String Theory}, Vol 1 and
2 (Cambridge University Press, Cambridge)

\bibitem{aa-lqg} Ashtekar A (2005) {\em New J. Phys.}, Vol. 7, p.
198 and references therein. \\ {\bf
http://stacks.iop.org/1367-2630/7/198}

\bibitem{smw-utr} Wagh S M (2005) {\em Foundations of a Universal
Theory of Relativity\/} and references therein  \\
{\bf http://arxiv.org/physics/0505063}

\bibitem{smw-let-0601} Wagh S M (2006) {\em Universal Theory of
Relativity and the ``Unification'' of Fundamental Physical
Interactions}, to be submitted.  {\bf
http://arxiv.org/physics/0602042}  \\ Wagh S M (2005) {\em
Progress with a Universal Theory of Relativity}, Talk delivered at
the SARS Einstein Centennial Meeting, Durban, September 25-26,
2005. \\ {\bf http://arxiv.org/physics/0602032} \\
Wagh S M (2005)  {\em Universal Relativity and Its Mathematical
Requirements}, Talk delivered at the SAMS 48th
Annual Meeting, Grahamstown, October 31 - November 2, 2005. \\
{\bf http://arxiv.org/physics/0602038}

\bibitem{maccallum} MacCallum M A H (2006) {\em Finding and using
exact solution of Einstein's equations}, {\bf
http://arxiv.org/gr-qc/0601102} \\ Kramer D, Stephani H, MacCallum
M A H and Herlt E (1980) {\it Exact Solutions of Einstein's Field
Equations} (Cambridge University Press, Cambridge) \\
C W Misner, K S Thorne and J A Wheeler (1973) {\em Gravitation} (W
H Freeman, New York)

\bibitem{chandra} Chandrasekhar S (1983) {\it Mathematical Theory of Black
Holes} (Clarendon Press, Oxford) \\ Penrose R (1998) in {\it Black
Holes and Singularities: S Chandrasekhar Symposium} (Ed. R M Wald,
Yale University Press, Yale) \\ Shapiro S and Teukolsky S A (1972)
{\it Black Holes, White Dwarfs and Neutron Stars} (Wiley
International, New York)

\bibitem{bhsupport} Broderick A E and Narayan R (2005) {\em On the
nature of compact dark mass of the galactic center}, {\bf
http://arxiv.org/astro-ph/0512211} and references therein

\bibitem{jvn01} Narlikar J V (2000) {\it A cosmic adventure}
(Rajhansa Prakashan, Pune, India) \\ Adler S L, Bazin C and
Schiffer R (1975) {\it Introduction to general relativity} (McGraw
Hill-Kogakusha, Tokyo) \\ Joshi P S (1993) {\it Global aspects in
gravitation and cosmology} (Clarendon Press, Oxford)

\bibitem{fhjvn} Hoyle F and Narlikar J V (1978) {\em Action at a
distance in Physics and Cosmology} (W H Freeman, New York)

\bibitem{qg-bang} Ashtekar A, Pawlowski T
\& Singh Parampreet (2005) {\em Quantum nature of the big bang},
{\bf http://arxiv.org/gr-qc/0602086}

\bibitem{pais} See also, Pais A (1982) {\it Subtle is the Lord ... The science
and the life of Albert Einstein} (Clarendon Press, Oxford)

\bibitem{pauli} Pauli W (1932) {\em Nature}, {\bf 20}, 186

\bibitem{ein-sci} Einstein A (1931) {\em Science}, {\bf 74}, p.
438

\bibitem{qgproblems} Ashtekar A and Statchel J (1991) {\it Conceptual
problems of quantum gravity} (Birkhauser, Boston) and references therein \\
Hawking S and Israel W Eds. (1979) {\it General Relativity - An
Einstein Centenary Survey} (Cambridge University Press, Cambridge)

\end{thebibliography}
\end{document}